\documentstyle[12pt]{article}
     \newcommand{\be}{\begin{equation}}
     \newcommand{\ee}{\end{equation}}
     \def\n{\noindent}
     \catcode `\@=11
     \catcode `\@=12

     \title{\bf\huge On Electrogravity Duality\footnote{Received Honorable
Mention in the 1998 Gravity Research Foundation Essay Competition}}
     
     \author{Naresh Dadhich\thanks{E-mail : nkd@iucaa.ernet.in} \\
     {\sl Inter-University Centre for Astronomy \& Astrophysics,}\\
     {\sl Post Bag 4, Ganeshkhind, Pune - 411 007, India.} 
     } 
     \date{}
     \begin{document}
     \maketitle
     
     \begin{abstract}
 
 \n By  resolving the gravitational field into electric and  magnetic 
 parts,  we  define an electrogravity duality  transformation  and 
 discover  an interesting  property of the field. Under  the 
 duality  transformation  a vacuum/flat spacetime  maps  into  the 
 original   spacetime   with  a  topological  defect   of   global 
 monopole/texture. The elctrogravity-duality is thus a topological 
 defect  generating  process.  It turns out  that  all  black  hole 
 solutions  possess  dual  solutions  that  imbibe  a  global 
 monopole. 
 
 \end{abstract}
 \newpage

 \n It  is well-known that gravitational field can, in  analogy  with 
 the electromagnetic field, be resolved into electric and magnetic 
 parts [1-5]. Electric part is caused  by  charge  (source) 
 while  magnetic part is due to motion of charge. In the  case  of 
 gravity  it would be mass-energy that would serve as charge  that 
 would  produce electric part and its motion would give  rise  to 
 magnetic part. Unlike other fields, gravitation has two kinds  of 
 charges,  the  one the usual matter-energy and the other  is  the 
 gravitational   field   energy.  They  will   lead   to   further 
 decomposition of electric part into active (the former, the usual 
 Coulombic)   and passive (the latter producing space  curvature). 
 For electromagnetic decomposition of gravitation, we resolve the
Riemann curvature into electric and magnetic parts reletive to a
timelike unit vector. Electric and magnetic parts are described by
second rank 3-tensors. Each of active and passive electric parts is
described by a symmetric tensor while magnetic part consists of a
symmetric Weyl magnetic part and an antisymmetric part representing energy
flux. 
 
 \n By elctrogravity duality we would mean interchange of  active 
 and passive electric parts of the field. It turns out that it 
 is  equivalent  to  interchange of the  Ricci  and  the  Einstein 
 curvatures. The vacuum equation would however remain invariant under 
 the duality transformation. The most interesting vacuum solutions 
 are  the black hole solutions and they are also unique.  In  obtaining 
 balck hole solutions, it turns out that one does not need to  use 
 all  the  equations;  i.e. there remains  one  equation free which  is 
 implied  by the others. For example in the simplest case  of  the 
 Schwarzschild solution the vacuum equation ultimately reduces  to 
 the  two equations, $\bigtriangledown^2 \phi = 0$ and $(r \phi)^{\prime}
= 0$ $g_{00}  =  - g_{11}^{-1}  = 1 + 2\phi(r)$, and a prime denotes
derivative w.r.t. $r$. Clearly the latter equation implies the  former. 
 That  is  even  if  we modify  the  vacuum  equation  by  putting 
 something  on the r.h.s. of the former, the equations will  still 
 yield  the  Schwarzschild  as  the  unique  solution.  Since  the 
 Schwarzschild solution is unique, the modified equation could  as 
 well  characterize  vacuum for spherical symmetry. The  modified 
 equation  would now  no  longer  be  duality-invariant.  The 
 interesting  question that arises is, what does the  solution  of 
 the dual set (dual-vacuum) represent? It turns out that the  dual 
 solution  represents the Schwarzschild black hole with  a  global 
 monopole charge [6]. Similarly it is possible to find solutions dual
to the Kerr and NUT solutions [7-8]. The duality transformation would thus
generate global monopole charge for  vacuum (including electrovac)
solutions in spherical and axial symmetries. \\

 \n As for vacuum, flat spacetime could as well be characterized by a
dulaity non-invariant eqaution. In the static case spacetime dual to flat
spacetime reperesents a massless pure monopole (zero mass limit of the
solution dual to the Schwarzschild black hole) while for the homogeneous
isotropic non-static case it is given by an FRW model with the equation of
state $\rho +  3 p  = 0$, which characterizes a global texture. \\

 \n  Global monopoles and  textures  are stable topological defects. They
are supposed to be produced when global symmetry is spontaneously broken
in phase transitions in the early Universe [9]. It is  remarkable 
 that  the  duality  transformation  automatically  incorporates  these 
 topological defects in the solutions of the Eistein equation in a 
 natural way. The elctrogravity-duality thus generates topological 
 defects  in  the vacuum/flat solutions of the Einstein equation.  This
is a remarkable new property.\\
 
 \n The  main  aim  of  this  essay is  to  expose  how  the  duality 
 transformation automatically includes the global monopole for the 
 stationary vacuum solutions and global texture for flat spacetime. We
shall demonstrate this for  the 
 Schwarzschild solution and flat spacetime. The same procedure would 
however work for the Kerr-NUT solutions. \\

  \n Let  us  begin  with  the resolution  of  the  Riemann  curvature 
     relative to a unit timelike vector, as follows :
     
     \be
     E_{ac} = R_{abcd} u^b u^d,  \tilde E_{ac} = *R*_{abcd} u^b u^d
     \ee
     
     \be
     H_{ac} = *R_{abcd} u^b u^d = H_{(ac)} - H_{[ac]}
     \ee
     
     \n where
     
     \be
     H_{(ac)} = *C_{abcd} u^b u^d
     \ee
     
     \be
     H_{[ac]} = \frac{1}{2} \eta_{abce} R^e_d u^b u^d.
     \ee
     
     \n Here $C_{abcd}$  is  the  Weyl conformal  curvature,
$\eta_{abcd}$
     is  the  4-dimensional volume element. $E_{ab} = E_{ba}, {\tilde E}_{ab}
     = {\tilde E}_{ba}, (E_{ab}, {\tilde E}_{ab}, H_{ab})
     u^b = 0,~ H= H^a_a = 0$ and $u^a u_a = 1$. 
     The Ricci tensor could then be written as
     
     \be
     R^b_a = E^b_a + {\tilde E}^b_a + (E + {\tilde E}) u_a u^b -
     {\tilde E } g^b_a + \frac{1}{2} (\eta_{amn} H^{mn} u^b + \eta^{bmn}
     H_{mn} u_a)
     \ee
     
     \n where $E = E^a_a$        and  $\tilde E = \tilde E^a_a$. 
     It may be noted that $E = (\tilde E + \frac{1}{2} T)/2$            defines 
     the  gravitational  charge density  while ${\tilde E}= - T_{ab}
     u^a u^b$            defines  the 
     energy density relative to the unit timelike vector $u^a$.   \\
    
    \n The vacuum equation, $R_{ab} = 0$ is in general equivalent to  
     
     \be
     E ~ or ~ {\tilde E} = 0,~ H_{[ab]} = 0 = E_{ab} + {\tilde E}_{ab}
     \ee
     
   \n which is symmetric in $E_{ab}$ and ${\tilde E}_{ab}$.

    \n We define the duality transformation as 
    
      \be
      E_{ab} \longleftrightarrow {\tilde E}_{ab}, ~H_{ab} = H_{ab}.
      \ee
    
    \n Thus the vacuum eqaution (6) is invariant under the duality 
    transformation (7). From eqn. (1) it is clear that the duality 
    transformation would map the Ricci tensor 
    into the Einstein tensor and vice-versa. This is because contraction of 
    Riemann is Ricci while of its double dual is Einstein.  \\   
    
    \n For obtaining the Schwarzschild solution, we consider the metric,
     
     \be
     ds^2 = c^2(r,t) dt^2 - a^2(r,t) dr^2 - r^2 (d \theta^2 + sin^2 \theta
     d \varphi^2).
     \ee
     
     \n The natural choice for the resolving vector in this case is of 
    course it being hypersurface orthogonal, pointing along the $t$-line. 
  From   eqn. (6), $H_{[ab]} = 0$ and $E^2_2 + {\tilde E}^2_2 = 0$ lead 
    to $ac = 1$ (for this, no boundary condition of asymptotic flatness need 
    be used). Now ${\tilde E} = 0$ gives $a = (1-2M/r)^{-1/2}$, which is the 
    Schwarzschild solution. Note that we did not need to use the remaining 
    equation  $E^1_1 + {\tilde E}^1_1 = 0$, it is  hence free and is
    implied by the rest. Without affecting the Schwarzschild solution, we 
    can introduce some distribution in the 1-direction. \\      
    
    \n We hence write the alternate equation as
     
     \be
     H_{[ab]} = 0 = {\tilde E},~ E_{ab} + {\tilde E}_{ab}
     = \lambda w_a w_b 
     \ee
     
     \n where $\lambda$    is a scalar, and $w_a$ is a spacelike unit
     vector parallel to the (radial) acceleration. It is clear that it
will also 
    admit the Schwarzschild solution as the general solution, and it determines
    $\lambda = 0$. That is for spherical symmetry the above alternate 
    equation also characterizes vacuum, because the Schwarzschild solution is 
    unique. \\
    
     \n  Let us now employ the duality  transformation (7) to the above  
    equation (9) to write  
     
     \be
     H_{[ab]} = 0 = E,~ E_{ab} + {\tilde E}_{ab} = \lambda w_a w_b.
     \ee
     
     \n Its general solution for the metric (8) is given by 
     
     \be
     c = a^{-1} = (1 - 2k - \frac{2M}{r})^{1/2}.
     \ee

     \n This is the Barriola-Vilenkin solution [6] for the  Schwarzschild 
     particle with global monopole charge, $\sqrt {2k}$. Again we shall 
       have $ac = 1$ and $E=0$ will then yield $c = (1-2k - 2M/r)^{1/2}$ 
       and $\lambda = 2k/r^2$. This  has  non-zero stresses given by
     
     \be
     T^0_0 = T^1_1 = \frac{2k}{r^2}.
     \ee

     \n A global  monopole    is    described    by    a    triplet    scalar,
      $\psi^a (r) = \eta f(r) x^a/r, x^a x^a = r^2$,
     which  through  the usual Lagrangian  generates  energy-momentum 
     distribution at large distance from the core precisely of the  form 
     given  above  in (12) [6].  Like the Schwarzschild solution the 
    monopole solution (11) is also the unique solution of eqn.(10). \\ 
  
     \n If we translate eqns. (9) and (10) in terms of the familiar Ricci 
  components, they would read as
  
     \be
     R^0_0 = R^1_1 = {\lambda},   R^2_2 = 0 = R_{01}
     \ee
  
     \n and
  
     \be
     R^0_0 = R^1_1 = 0 = R_{01},  R^2_2 = {\lambda}.
      \ee
  
      \n For the metric (8), eqns. $R^0_0 = R^1_1 = R_{01} = 0$ lead to 
       $ac = 1$ and $c^2 = f(r) =1 + 2 \phi$, say, and then  
 
      \be
      R^0_0 = - \bigtriangledown^2 \phi 
      \ee
 
      \be
      R^2_2 = - \frac {2}{r^2} (r \phi)^{\prime}.
      \ee 
 
     \n Now the set (13) integrates to give ${\phi} = - M/r$ and 
  ${\lambda} = 0$, which is the Schwarzschild solution while (14) will 
  give ${\phi} = - k - M/r$ and ${\lambda} = 2k/{r}^2$, the 
  Schwarzschild with global monopole charge. Thus global monopole owes 
  its existence to the constant $k$, appearing in the solution of the 
  usual Laplace equation implied by eqns. (14) and (15). It defines a 
  pure gauge for the Newtonian theory, which could be chosen freely, 
  while the Einstein vacuum equation determines it to be zero. For the 
  dual-vacuum equation (14), it is free like the Newtonian case but it 
  produces non-zero curvature and hence would represent non-trivial 
  physical and dynamical effect (see $R^2_2 = - 2k/{r}^2 \neq 0$ 
  unless $k = 0$). This is the crucial difference between the 
  Newtonian theory and GR in relation to this problem, that the latter 
  determines the relativistic potential ${\phi}$ absolutely, vanishing 
  only at infinity. This freedom is restored in the dual-vacuum equation, 
  of course at the cost of introducing stresses that represent a global 
  monopole charge. The uniform potential would hence represent a massless 
  global monopole ($M = 0$ in the solution (11)), which is solely 
  supported by the passive part of electric field. It has been argued and 
  demonstrated [5] that it is the non-linear aspect of the field (which 
  incorporates interaction of gravitational field energy density) that 
  produces space-curvatures and consequently the passive electric part. It 
  is important to note that the relativistic potential ${\phi}$ plays the 
  dual role of the Newtonian potential as well as the non-Newtonian role 
  of producing curvature in space. The latter aspect persisits even when 
  potential is constant different from zero. It is the dual-vacuum 
  equation that uncovers this aspect of the field. \\
    
      \n On the other hand, flat spacetime could also in alternative form be 
    characterized by 
     \be
     {\tilde E}_{ab} = 0 = H_{[ab]}, E_{ab} = \lambda w_a w_b
     \ee
     
     \n leading to $c=a=1$, and implying ${\lambda} = 0$ . Its dual will be 
     
     \be
     E_{ab} = 0 = H_{[ab]}, {\tilde E}_{ab} = \lambda w_a w_b
     \ee

     \n yielding the general solution,
     
     \be
     c^{\prime} = a^{\prime} = 0 \Longrightarrow c=1, a = const. = (1-2k)^{-1/2}
     \ee

     \n which is non-flat and represents a global monopole of zero  mass, 
     as it follows from the solution (11) when $M=0$. This is also the 
  uniform relativistic potential solution. \\
   
     \n Further it  turns  out that a perfect fluid with $\rho + 3p = 0$ 
     goes  to 
     flat spacetime under the duality transformation [3]. This is  the 
     equation  of  state,  which means $E=0$, that characterizes global 
     texture  [7,19]. That  is, the necessary condition  for spacetimes of 
    topological defects;  global  textures  and monopoles is   $E = 0$. The  
    non-static homogeneous and isotropic (here $w_a$ is an isotropic unit
    vector) solution of the dual-flat equation (18) is the FRW 
    metric with $\rho +3p = 0$, which  
    determines  the scale factor $S(t) = \alpha t + \beta, $ and $\rho = 3
     (\alpha^2 + n) / (\alpha t + \beta)^2, ~ n = \pm 1, 0$. This is also 
    the unique solution. The general solutions of the 
    dual-flat equation are thus the massless global monopole (uniform 
    potential) spacetime in the static case and the global texture spacetime 
    in the non-static homogeneous and isotropic  case. Thus they are dual
     to flat spacetime. \\

  \n The  above  method would straightway work for the  charged  black 
 hole  as well as for the de Sitter spacetime. In the latter  case 
 the  duality only changes the sign of the cosmological  constant, 
 i.e.  de  Sitter to anti de Sitter. Though the  calculations  are 
 considerably  more involved for the Kerr and NUT  solutions,  
 the  prescription works and the solutions dual to them have  been 
 found [7-8]. We have thus constructed electrogravity-dual  solutions 
 to  all  the black hole solutions including the NUT  solution  as 
 well  as to flat spacetime representing massless global  monopole 
 and global texture. \\ 
 
 \n This procedure of constructing dual solutions would work so  long 
 as there occurs a free equation in the Einstein set which is  not 
 used  in finding the solution. Note that this is so for  all  the 
 black  hole  solutions.  Then the dual  set  admits  a  
 solution  like the original set which incorporates a  topological 
 defect in the original spacetime. This is a remarkable property   of   
  the  gravitational   field, which is exposed by the duality
transformation. The duality 
 transformation for a fluid source has been considered [10], and  it
turns out that  a fluid solution maps into a fluid  solution  with 
 $\rho \leftrightarrow (\rho + 3 p)/2 , p \leftrightarrow (\rho  - 
 p)/2$, and heat flux and pressure anisotropy remaining unaltered. 
 This shows that the stiff fluid is dual to dust, $\rho + 3 p  =0$ 
 is as expected dual to flat spacetime, and the radiation and  the 
 de Sitter models are self dual.\\
 
 \n It could be argued that duality-related spacetimes do share the
basic gravitational property at the Newtonain level. Could the duality
transformation hence signify "maximal closeness" or "minimal
difference" between them? Then dual-flat spacetimes would be
``minimally'' curved. By modifying the vacuum/flat equation we have been 
able to construct solutions dual to black holes and flat spacetime. The
duality transformation only produces a topological defect in terms of
global monopole/texture in the original spacetime. This appears to be an
interesting property of the field. What good it is in terms of its
applications remains to be investigated. At any rate it is a novel way
of creating topological defects in the vacuum/flat spacetimes. \\

\n  Acknowledgement: I thank L.K. Patel for fruitful collaboration for
finding the dual-Kerr solution and Mohammad Nouri-Zonoz and Donald
Lynden-Bell for the dual-NUT solution. \\

\end{document}